\pdfoutput=1
\documentclass[a4paper, 12pt]{article}
\usepackage{NOTES-API}
\usepackage[style=apa]{biblatex} 
\usepackage[title]{appendix}
\myname{Zhengdao LI, Penggao Yan, Baoshan Song and Li-Ta Hsu}
\title{\textbf{Improved GNSS Positioning in Urban Environments Using a Logistic Error Model}}

\begin{document}
\maketitle
 
\section*{Abstract}
A Gaussian error assumption is commonly adopted in the pseudorange measurement model for global navigation satellite system (GNSS) positioning, which leads to the conventional least squares (LS) estimator. In urban environments, however, multipath and non-line-of-sight (NLOS) receptions produce heavy-tailed pseudorange errors that are not well represented by the Gaussian model. This study models urban GNSS pseudorange errors using a logistic distribution and derives the corresponding maximum likelihood estimator, termed the Least Quasi-Log-Cosh (LQLC) estimator. The resulting estimation problem is solved efficiently using an iteratively reweighted least squares (IRLS) algorithm. Experiments in light, medium, and deep urban environments show that LQLC consistently outperforms LS, reducing the three-dimensional (3D) root mean square error (RMSE) by approximately 11\%--31\% and the 3D error standard deviation (STD) by approximately 27\%--61\%. A controlled scale-mismatch analysis further shows that LQLC is more sensitive to severe underestimation than to overestimation of the logistic scale, indicating that the practical tuning requirement is to avoid overly small scale values rather than to achieve exact scale matching. In addition, the computational cost remains compatible with real-time positioning. These results indicate that logistic modeling provides a simple and practical alternative to Gaussian-based urban GNSS positioning.

\section{Introduction}
Global navigation satellite system (GNSS) positioning is widely used in transportation and location-based services. For pseudorange-based positioning, the conventional formulation assumes Gaussian measurement errors, under which the maximum likelihood estimator (MLE) reduces to the well-known least squares (LS) estimator \parencite{kaplan_understanding_2006}. However, this assumption becomes questionable in urban environments, where high-rise buildings and dense infrastructure frequently block, reflect, and diffract satellite signals \parencite{Groves_2011,Hsu_2017,Lee_2022}. The resulting multipath and non-line-of-sight (NLOS) receptions can produce large pseudorange errors more frequently than the Gaussian model predicts and can lead to heavy-tailed error behavior \parencite{crespillo_design_2020,Weng_2023,peretic_statistical_2025}. Consequently, the statistical basis of Gaussian-based positioning is weakened in urban scenarios, and this issue can affect both position estimation and integrity monitoring \parencite{medina_robust_2019a,zhu_gnss_2018,yan_highavailability_2025}. \par
To tackle this issue, urban GNSS positioning research has explored a broad range of mitigation routes. Representative directions include shadow or sidewalk matching with environmental context \parencite{Groves_2011,Weng_2025}, 3D-building-, light detection and ranging (LiDAR)-, or vision-aided NLOS mitigation \parencite{Hsu_2017,Wen_2022,wen_vision_2023}, tightly coupled inertial or precise point positioning (PPP) integration for challenging urban conditions \parencite{yoder_inertial_2023,vana_ppp_2023,fu_android_2026}, and pseudorange-bias estimation or multipath prediction \parencite{Marais_2013,Lee_2022,khanafseh_gnss_2018a}. Integrity- or optimization-oriented urban positioning formulations have also been investigated \parencite{Wen_2021,Gupta_2024,xia_integrity_2024,yan_multiple_2025}. Although these approaches can improve positioning performance, many of them rely on additional environmental information, dedicated auxiliary sensing, or more elaborate processing pipelines. \par
From a statistical-modeling perspective, previous studies have also considered replacing the Gaussian error model itself. Gaussian mixture models (GMMs) have been used to represent multimodal or fault-contaminated GNSS error behavior \parencite{blanch_position_2008,Gupta_2024, yan_principal_overbound_2025} and the Student's-t \parencite{dhital_new_2013,tao_improved_2025} and Cauchy distributions \parencite{li_paired_overbound_2025}, have been adopted to accommodate heavier tails within GNSS estimation. Although these models can offer improved empirical fit, they are less attractive when the objective is to obtain a positioning estimator that is both statistically better matched and straightforward to implement.\par
The logistic distribution is a promising candidate in this context. Like the Gaussian model, it is characterized by a location parameter and a scale parameter, but it has heavier tails and therefore assigns higher probability to large residuals \parencite{johnson_continuous_1995}. Existing studies have reported encouraging empirical results for logistic modeling in challenging GNSS environments. El-Mowafy showed that the logistic distribution provided a better fit than the Gaussian model for a large sample of urban position errors \parencite{el-mowafy_fault_2020}. The European Space Agency's INSPIRe project similarly found the logistic distribution to be the best overall fit among several commonly used statistical models for pseudorange errors in crowd-sourced GNSS datasets \parencite{_assess_2024}. However, the potential of a logistic error assumption for GNSS position estimation remains insufficiently explored.\par
In this work, we address this gap by deriving a logistic-based MLE for the linearized pseudorange model. The resulting estimator, termed the Least Quasi-Log-Cosh (LQLC) estimator, belongs to the class of M-estimators \parencite{huber_robust_1981}. Owing to the explicit form of the logistic probability density function, the associated objective function can be written in closed form and solved efficiently using iteratively reweighted least squares (IRLS) \parencite{huber_robust_1964,maronna_robust_2019}. The proposed method is evaluated using real GNSS data collected in light, medium, and deep urban environments in Hong Kong with Global Positioning System (GPS) and Beidou L1 observations from a u-blox F9P receiver. A controlled scale-mismatch analysis is further conducted to clarify the practical effect of logistic-scale mismatch. The contributions of this study are threefold:
\begin{itemize}
    \item A logistic-based MLE, the LQLC estimator, is derived for urban GNSS positioning, along with an efficient IRLS implementation.
    \item The proposed estimator is validated experimentally in representative urban environments using real GNSS observations.
    \item A controlled scale-mismatch analysis is introduced to examine the practical sensitivity of LQLC to imperfect logistic-scale fitting.
\end{itemize}
The remainder of this paper is organized as follows. Section \ref{sec:distribution} examines the distributional properties of urban pseudorange errors. Section \ref{sec:lqlc} presents the LQLC estimator and its IRLS solution. Section \ref{sec:experiment_results} reports the real-data evaluation, the computation-efficiency analysis, and the controlled scale-mismatch study. Finally, Section \ref{sec:conclusion} concludes the paper.

\section{Distributional properties of urban pseudorange errors}\label{sec:distribution}

This section examines the distributional characteristics of pseudorange errors in three representative urban environments, namely light, medium, and deep urban scenarios. The purpose is to assess whether the Gaussian model commonly adopted in GNSS positioning remains adequate under these conditions and, if not, to identify a simple alternative that better reflects the observed error behavior. To this end, the empirical error distributions are compared with four candidate statistical models: Gaussian, logistic, bimodal Gaussian mixture model (BGMM), and Student's t.\par

\begin{figure}[htb!]
    \centering
    \addsubFig{0.31}{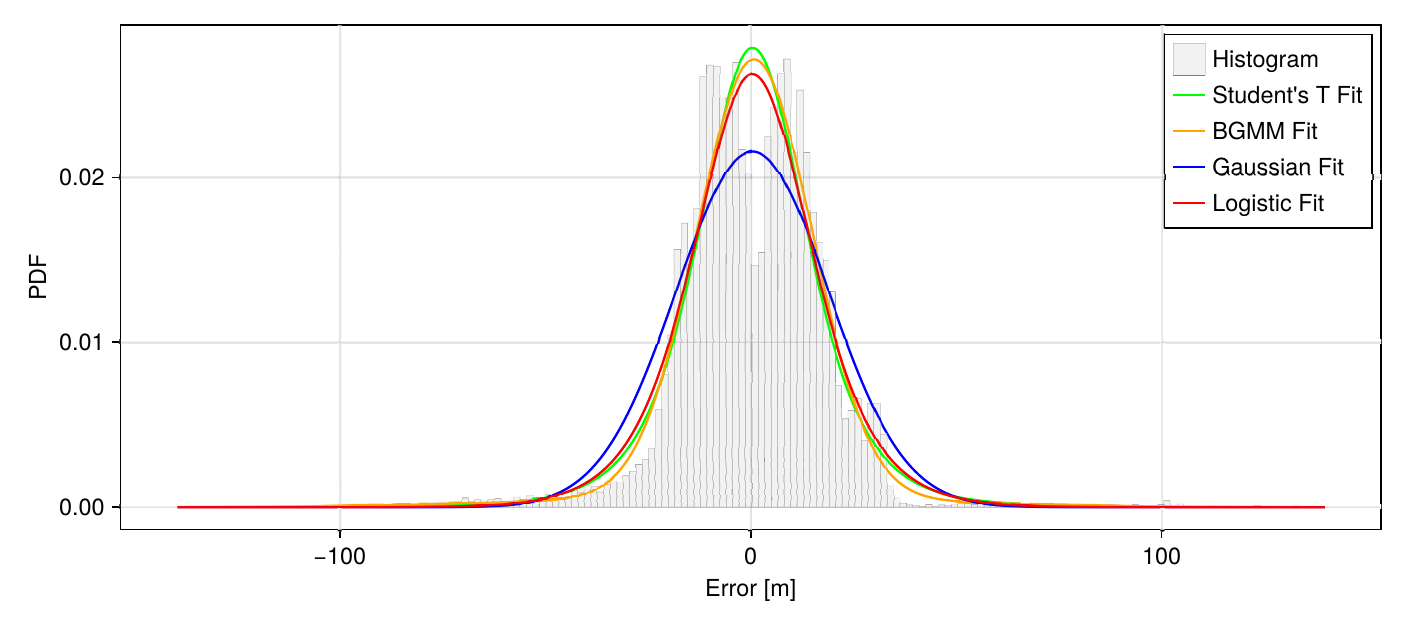}{}
    \addsubFig{0.31}{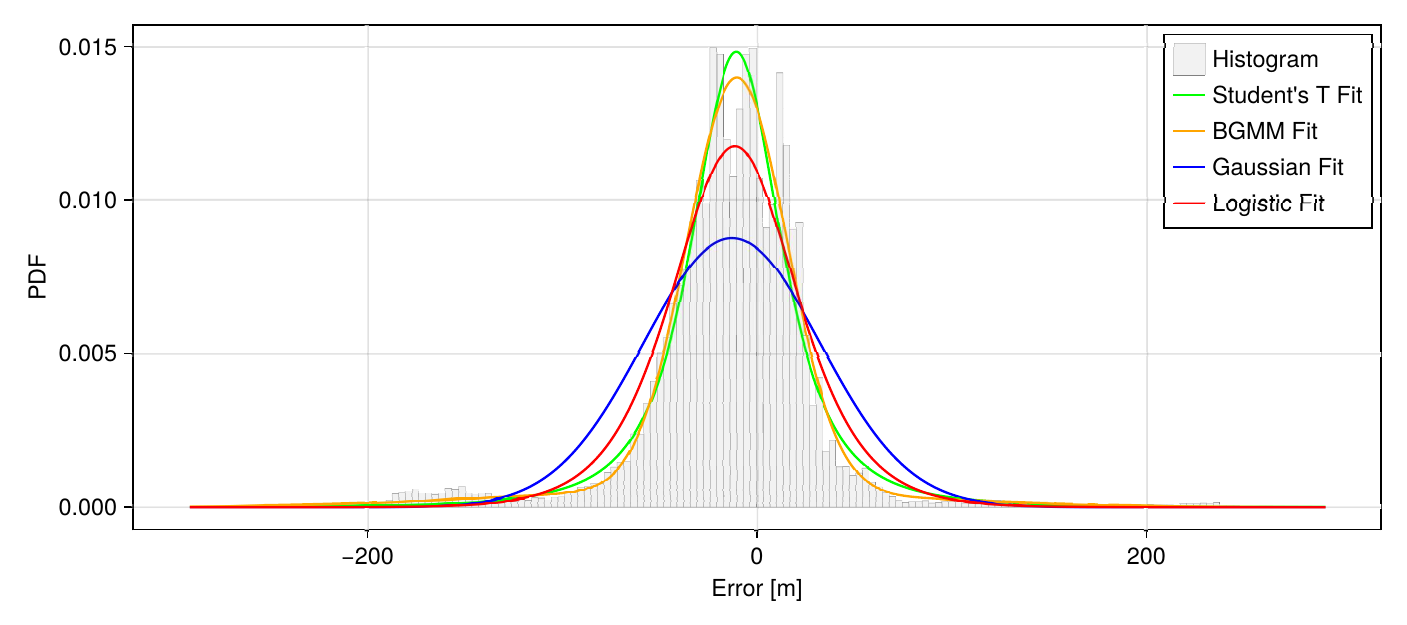}{}
    \addsubFig{0.31}{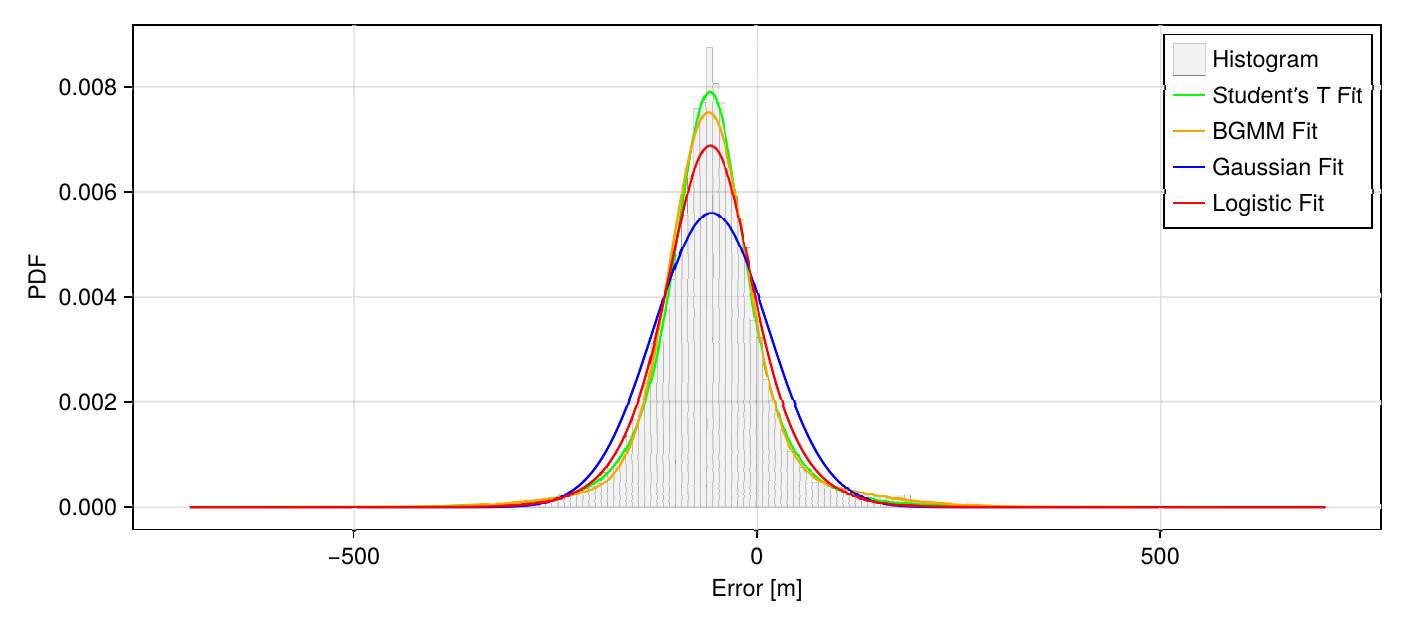}{}
    \caption{Fitted distributions for the urban pseudorange error datasets: (a) light urban; (b) medium urban; and (c) deep urban.}
    \label{fig: urban fitted distributions}
\end{figure}

The data used here were collected in Hong Kong and are the same datasets later used for the urban positioning evaluation. At this stage, they are introduced only for distribution fitting; the full data-acquisition procedure, positioning setup, and evaluation details are provided in Section \ref{sec:experiment_results}. Within each urban dataset, the pseudorange measurements from all visible satellites are pooled into a single sample and fitted using the candidate distributions. After pooling, the light, medium, and deep urban datasets contain 51,837, 57,682, and 50,517 pseudorange error samples, respectively. For the BGMM, the model parameters are estimated using the Expectation-Maximization (EM) algorithm. For the Gaussian, logistic, and Student's t distributions, parameters are obtained through standard maximum likelihood estimation. As a result, each dataset is represented by one fitted parameter set for each candidate model. Although fitting by satellite elevation interval is common in larger datasets, it was not adopted here because the available urban data would become too sparse within each bin to support stable and reliable fitting. Pooling all measurements within each dataset therefore provides a more defensible basis for distribution comparison in the present study. The fitted parameters for the three urban datasets are summarized jointly in Table \ref{tab: urban fitted dist parameters}, and the corresponding fitted curves are shown together in Figure \ref{fig: urban fitted distributions}.

\begin{table}[htb]
 \caption{Fitted distribution parameters for the light, medium, and deep urban datasets.}
 \label{tab: urban fitted dist parameters}
\begin{tblr}{colspec={X[c]X[c]X[c]X[c]},
width=\textwidth,
row{even} = {white,font=\small},
row{odd} = {bg=black!10,font=\small},
row{1} = {bg=black!20,font=\bfseries\small},
hline{Z} = {1pt,solid,black!60},
rowsep=3pt
}
Fitted distribution & Light urban & Medium urban & Deep urban  \\
Gaussian &
\makecell[c]{$(\mu,\sigma)$ \\ $(0.33\text{m},18.49\text{m})$} &
\makecell[c]{$(\mu,\sigma)$ \\ $(-12.95\text{m},45.47\text{m})$} &
\makecell[c]{$(\mu,\sigma)$ \\ $(-56.91\text{m},71.36\text{m})$} \\
BGMM &
\makecell[c]{$p_1=0.93$ \\ $\mu_1=0.82\text{m}$ \\ $\sigma_1=13.97\text{m}$ \\ $\mu_2=-6.25\text{m}$ \\ $\sigma_2=47.72\text{m}$} &
\makecell[c]{$p_1=0.14$ \\ $\mu_1=-30.00\text{m}$ \\ $\sigma_1=103.51\text{m}$ \\ $\mu_2=-10.27\text{m}$ \\ $\sigma_2=25.58\text{m}$} &
\makecell[c]{$p_1=0.79$ \\ $\mu_1=-60.83\text{m}$ \\ $\sigma_1=46.08\text{m}$ \\ $\mu_2=-41.71\text{m}$ \\ $\sigma_2=127.68\text{m}$} \\
Student's t &
\makecell[c]{$(c,\lambda,\nu)$ \\ $(0.46\text{m},13.58\text{m},4.68)$} &
\makecell[c]{$(c,\lambda,\nu)$ \\ $(-10.64\text{m},24.40\text{m},2.50)$} &
\makecell[c]{$(c,\lambda,\nu)$ \\ $(-59.00\text{m},46.74\text{m},3.18)$} \\
Logistic &
\makecell[c]{$(m,s)$ \\ $(0.42\text{m},9.52\text{m})$} &
\makecell[c]{$(m,s)$ \\ $(-11.66\text{m},21.28\text{m})$} &
\makecell[c]{$(m,s)$ \\ $(-58.54\text{m},36.33\text{m})$} \\
\end{tblr}
\end{table}

Figure \ref{fig: urban fitted distributions} shows that the Gaussian model already departs visibly from the empirical tails in the light urban dataset, whereas the logistic, BGMM, and Student's t models provide more plausible fits. This discrepancy becomes more evident in the medium urban dataset and is most pronounced in the deep urban dataset, where the Gaussian model is clearly inadequate for representing the observed error distribution. Across all three scenarios, the urban pseudorange errors therefore exhibit clear heavy-tailed characteristics.\par

Among the candidate models, BGMM and Student's t provide flexible descriptions of the empirical distributions, particularly in the tails. However, this flexibility comes at the expense of increased model complexity or a less convenient estimation structure for positioning. The logistic distribution offers a more practical compromise. It captures the heavy-tailed behavior more faithfully than the Gaussian model while retaining a simple location-scale parameterization similar to the Gaussian case. For the purpose of estimator design, this combination of empirical adequacy and model simplicity makes the logistic distribution the most suitable choice in the present study.\par

These observations motivate the estimator developed in the next section. Since the logistic distribution provides a more realistic yet still tractable description of urban pseudorange errors, it is natural to derive the GNSS positioning estimator directly from a logistic error assumption rather than from the conventional Gaussian model.

\section{Least Quasi-Log-Cosh (LQLC) Estimator}\label{sec:lqlc}

\subsection{Basis of M-estimator for GNSS positioning}
For $n$ satellites in view, the linearized GNSS pseudorange measurement model is written as
\begin{equation}
    \mathbf{y}=\mathbf{H}\mathbf{x}+ \bm{\varepsilon}, \label{equ: pseudo mea model}
\end{equation}
where \textbf{y} is the vector of differences between the measured and predicted pseudoranges, $\textbf{H}$ is the geometry matrix, $\textbf{x}$ is the state vector containing the user position and clock bias, and $\bm{\varepsilon}$ is the measurement error vector.\par

Under a specified probabilistic model for the measurement errors, the unknown state vector $\textbf{x}$ can be estimated by maximum likelihood estimation (MLE). The resulting estimator belongs to the general class of M-estimators \parencite{huber_robust_1964}, which can be written as
\begin{equation}
     \hat{\textbf{x}}=\argmin \sum_{i=1}^n \mathcal{J}_i \prt{\frac{\textbf{y}_{(i)}-\textbf{H}_{(i,:)}\textbf{x}}{\nu_i}} =  \argmin \sum_{i=1}^n  \mathcal{J}_i \prt{\bar{r}_i}   \label{equ: general form of M-estimator},
\end{equation}
where $\mathcal{J}_i$ is the loss function, $\nu_i$ is the scale parameter associated with the $i$th observation, and $\bar{r}_i$ is the normalized residual \parencite{crespillo_design_2020}. The subscript $(i)$ denotes the $i$th element of a vector, whereas $(i,:)$ denotes the $i$th row of a matrix. This notation is used throughout the paper.

\subsection{Construction of LQLC estimator}

Motivated by the heavy-tailed empirical behavior shown in the previous section, each measurement error is modeled using the logistic distribution. Its probability density function (PDF) $f_L$ and cumulative distribution function (CDF) $F_L$, with location parameter $m$ and scale parameter $s$, are given by
\begin{align}
    f_L(x; m, s) &= \frac{e^{-\frac{x-m}{s}}}{s \prt{1+e^{-\frac{x-m}{s}}}^2} = \frac{1}{s  \prt{e^{\frac{x-m}{s}}+e^{-\frac{x-m}{s}} +2}}, \label{equ: logi pdf} \\
    F_L(x; c, s) &= \frac{1}{1+e^{-\frac{x-m}{s}}} \label{equ: logi cdf}.
\end{align}\par

If the individual measurement errors are assumed to follow a zero-mean logistic distribution,
\begin{equation}
      \bm{\varepsilon}_{(i)} \sim  \text{L}(0, s_i),\quad \forall i\in[1, n], \label{equ: error follows logi}
\end{equation}
then maximizing the likelihood is equivalent to minimizing the negative log-likelihood, which yields
\begin{equation}
    \hat{\textbf{x}}=\argmin \sum_{i=1}^n \ln \prt{\cosh \prt{ \frac{\textbf{y}_{(i)}-\textbf{H}_{(i,:)} \textbf{x}}{s_i}}+1}=\argmin  \sum_{i=1}^n \mathcal{J}_{QLC,i},
    \label{equ: LQLC estimator}
\end{equation}
where $\mathcal{J}_{QLC,i}$ denotes the quasi-log-cosh (QLC) loss for the $i$th observation. We refer to the resulting estimator as the Least Quasi-Log-Cosh (LQLC) estimator. The corresponding MLE derivation is summarized in Appendix \ref{app: logi mle}. 

\subsection{Numerical solution for LQLC estimator}

\subsubsection{Comparison between the LS and LQLC estimators}

\begin{figure}[htb!]
    \centering
    \includegraphics[width=\textwidth]{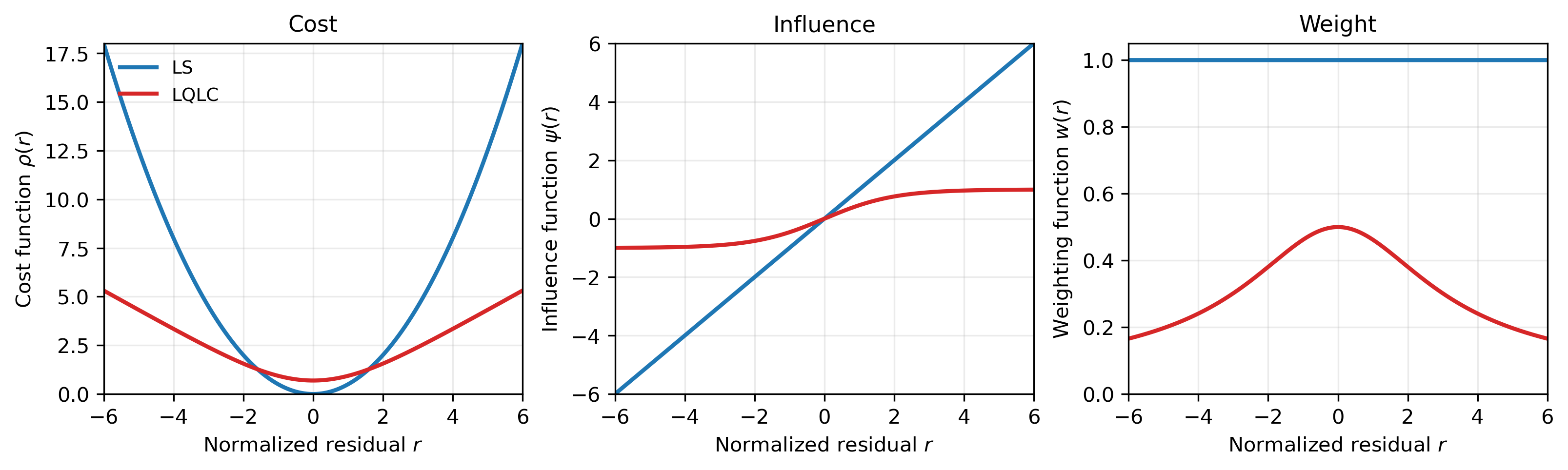}
    \caption{Visual comparison of the cost, influence, and weighting functions induced by the Gaussian and logistic error models. The LQLC estimator retains a smooth convex cost, while its influence and weighting functions become progressively bounded for large normalized residuals.}
    \label{fig: ls lqlc cost influence weight}
\end{figure}

Table \ref{tab: comparing cost/influence/weighting} and Figure \ref{fig: ls lqlc cost influence weight} summarize the M-estimator forms induced by the Gaussian and logistic error assumptions. Under the Gaussian model, the cost function is quadratic, the influence function increases linearly with the residual, and the induced weighting function remains constant. This leads to the conventional LS estimator. Under the logistic model, however, the cost function becomes quasi-log-cosh, the influence function is bounded by the hyperbolic tangent term, and the corresponding weight decreases with increasing residual magnitude. Therefore, the difference between LS and LQLC is not only a change of likelihood model, but also a change in how large residuals are treated during estimation. This comparison also explains why LQLC can still be solved within an iteratively reweighted least squares framework, which is described next.\par

\begin{table}[htb!]
    \caption{Cost, influence, and weighting functions of the M-estimators induced by Gaussian and logistic error assumptions.}
    \label{tab: comparing cost/influence/weighting}
    \begin{tblr}
            {colspec={X[c]X[c]X[c]X[c]X[c]},
            width=\textwidth,
            row{even} = {white,font=\small},
            row{odd} = {bg=black!10,font=\small},
            row{1} = {bg=black!20,font=\bfseries\small},
            hline{Z} = {1pt,solid,black!60},
            rowsep=3pt
            }
            \textbf{Assumed error model} & \textbf{Estimator given by MLE}& \textbf{Cost function ($\rho(r)$)} & \textbf{Influence function ($\psi(r)$)} & \textbf{Weighting function ($w(r)$)}  \\
            Gaussian& LS & $\frac{1}{2}r^2$ & $r$ & $1$ \\
            Logistic&  LQLC &$\ln \left(\cosh(r)+1\right)$ & $\tanh \prt{\frac{r}{2}}$ & $\frac{\tanh \prt{\frac{r}{2}}}{r}$ \\
    \end{tblr}
\end{table}

\subsubsection{IRLS solver for LQLC estimator}
As indicated by the residual-dependent weighting structure in Figure \ref{fig: ls lqlc cost influence weight} and Table \ref{tab: comparing cost/influence/weighting}, the proposed LQLC estimator can be solved efficiently by iterative reweighted least squares (IRLS) \parencite{maronna_robust_2019}. The key derivation is summarized below.\par

Starting from Equation \eqref{equ: general form of M-estimator}, a necessary optimality condition is obtained by setting the first derivative with respect to $\textbf{x}$ equal to zero:
\begin{equation}
    \sum_{i=1}^n \pd{\mathcal{J}_{i}(\bar{r}_i)}{\textbf{x}}=\sum_{i=1}^n \prt{\pd{\bar{r}_i}{\textbf{x}}}^\top\pd{\mathcal{J}_{i}(\bar{r}_i)}{\bar{r}_i}=0.  \label{equ: irls first derivative main}
\end{equation}
By introducing the influence function $\psi_i(\bar{r}_i)=\frac{\partial \mathcal{J}_i(\bar{r}_i)}{\partial \bar{r}_i}$, Equation \eqref{equ: irls first derivative main} can be rewritten as
\begin{equation}
    \sum_{i=1}^n \prt{\pd{\bar{r}_i}{\textbf{x}}}^\top \bar{r}_i \cdot \prt{\frac{\psi_i(\bar{r}_i)}{\bar{r}_i}} = 0.
\end{equation}
If the ratio $\frac{\psi_i(\bar{r}_i)}{\bar{r}_i}$ is treated as a constant computed from the residuals of the previous iteration, then the problem is converted into a weighted least-squares form. Defining
\begin{equation}
    w_i=\frac{\psi_i(\bar{r}_i)}{\bar{r}_i},
\end{equation}
the corresponding iterative subproblem becomes
\begin{equation}
    \hat{\textbf{x}}=\argmin\sum_{i=1}^n w_i \prt{\bar{r}_i(\textbf{x})}^2. \label{equ: weighted sum of squares main}
\end{equation}
For the normalized residual $\bar{r}_i=\frac{\textbf{y}_{(i)}-\textbf{H}_{(i,:)}\textbf{x}}{\nu_i}$, Equation \eqref{equ: weighted sum of squares main} yields the weighted normal equation
\begin{equation}
    \textbf{H}^\top \textbf{W}\prt{\textbf{y}-\textbf{H}\textbf{x}} = 0,
\end{equation}
where $\textbf{W}$ is diagonal with entries $W_{(i,i)}=\frac{w_i}{\nu_i^2}$. Therefore, each IRLS iteration updates the state estimate according to
\begin{equation}
    \hat{\textbf{x}}=\prt{\textbf{H}^\top \textbf{W}\textbf{H}}^{-1}\textbf{H}^\top \textbf{W}\textbf{y}. \label{equ: irls solution main}
\end{equation}

For the LQLC estimator, the loss function is $\mathcal{J}_{QLC}(r)=\ln(\cosh(r)+1)$, and the corresponding influence and weighting functions are
\begin{equation}
    \psi_{QLC}(r)=\tanh\prt{\frac{r}{2}}, \qquad
    w_{QLC}(r)=\frac{\tanh\prt{\frac{r}{2}}}{r}.
\end{equation}
Hence, large normalized residuals are assigned smaller weights automatically, which gives the estimator its robustness to multipath-contaminated observations. Because the QLC loss is convex, the IRLS procedure converges to the unique minimizer of the objective. The corresponding implementation is summarized in Algorithm \ref{alg: LQLC IRLS}.

\begin{algorithm}[!ht]
\caption{LQLC estimation using IRLS}
\label{alg: LQLC IRLS}
\begin{algorithmic}[1]
    \State Initialize $\textbf{W=I}$
    \While{no convergence is reached}
        \State Compute the vector of delta pseudorange $\textbf{y}$, and observation matrix $\textbf{H}$
        \State Compute weighted least-squares (WLS) solution $\hat{\textbf{x}}=\prt{\textbf{H}^\top \textbf{WH}}^{-1}\textbf{H}^\top \textbf{W}\textbf{y}$
        \State Compute each residual $r_i=\frac{\textbf{y}_{(i)}-\textbf{H}_{(i,:)}\hat{\textbf{x}}}{s_i}$
        \State $\textbf{W}_{(i,i)} \gets \frac{\tanh\prt{\frac{r_{(i)}}{2}}}{s_i^2 r_{(i)}}$
    \EndWhile
\end{algorithmic}
\end{algorithm}

\section{Experiment and results}\label{sec:experiment_results}

\subsection{Experiment setting}\label{sec:experiment_setting}
The proposed method was evaluated using the same three Hong Kong urban datasets in Section \ref{sec:distribution}. The datasets were collected in light, medium, and deep urban scenarios, which represent increasing levels of signal obstruction and multipath contamination. For each dataset, a u-blox F9P receiver was used to track GPS and Beidou L1 signals, and a high-precision SPAN-CPT reference system provides the ground truth.  

For each dataset, the Gaussian and logistic parameter sets obtained from the pooled distribution fitting in Section \ref{sec:distribution} were used in the corresponding LS and LQLC based single point positioning (SPP) solutions. The resulting positioning performance was compared using the three-dimensional root mean square error (RMSE) and the three-dimensional error standard deviation (STD). All positioning algorithms were implemented and tested on a PC equipped with an Intel Core i7-1355U processor.\par

\subsection{Positioning results comparison}

\subsubsection{Light urban}

\begin{figure}[htb!]
    \addsubFig{0.44}{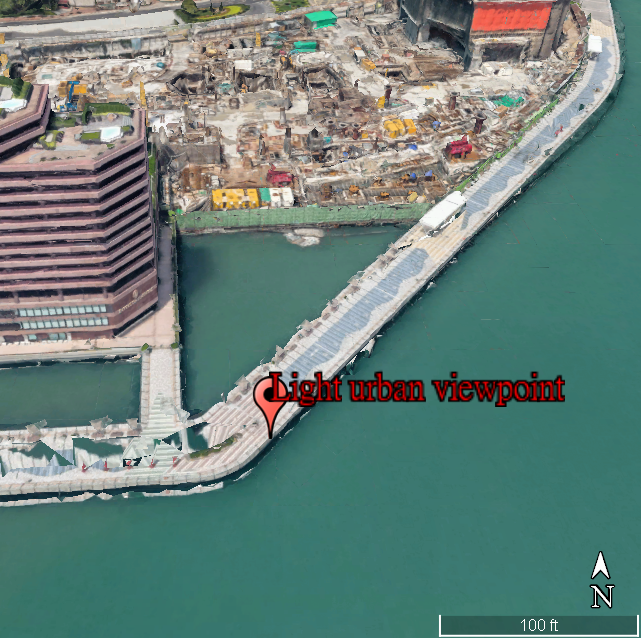}{}
    \addsubFig{0.49}{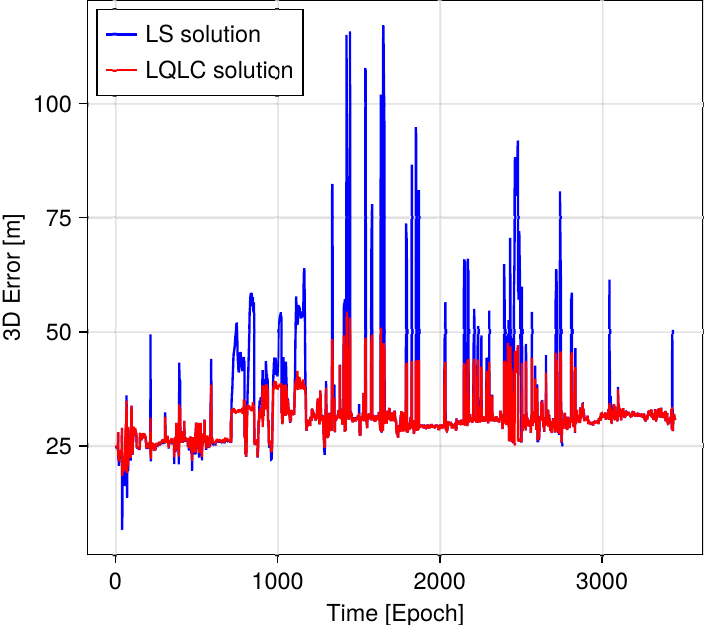}{}
    \caption{Analysis of the light urban dataset collected at the latitude, longitude, and height (LLH) position of (22.293079, 114.17489, 3.0m): (a) Google Earth snapshot of the receiver's surroundings; (b) 3D positioning errors given by LS and LQLC estimators}
    \label{fig: light urban}
\end{figure}

Figure \ref{fig: light urban}a shows that the light urban test site is located on a waterfront promenade, with open sky on one side and tall buildings and construction works on the other. This geometry creates a mixed propagation environment in which direct and reflected signals coexist, producing a representative but comparatively moderate urban positioning challenge.\par 

\begin{table}[htb!]
    \caption{Comparison between LS and LQLC positioning solutions in the light urban environment.}
    \label{tab: light urban}
    \begin{tblr}
            {colspec={X[c]X[c]X[c]X[c]},
            width=\textwidth,
            row{even} = {white,font=\small},
            row{odd} = {bg=black!10,font=\small},
            row{1} = {bg=black!20,font=\bfseries\small},
            hline{Z} = {1pt,solid,black!60},
            rowsep=3pt
            }
            \textbf{Metric}& \textbf{LS solutions} & \textbf{LQLC solutions} & \textbf{Percentage of reduction}  \\
            3D RMSE & 34.82m & 30.96m & 11.06\% \\
            3D STD & 11.13m & 4.45m & 59.99\%
    \end{tblr}
\end{table}

Figure \ref{fig: light urban}b shows the 3D positioning errors of LS and LQLC over the observation period. The LS solution exhibits clear instability, with repeated error spikes between approximately 60\,m and 120\,m, especially between epochs 1200 and 2700. By contrast, the LQLC solution varies more moderately and suppresses the largest excursions, with a maximum error below 55\,m.\par

In single-point positioning, such spikes are typically associated with multipath-contaminated pseudorange measurements. Because LS is derived under a Gaussian error model, it remains sensitive to large residuals that are assigned very low probability under that assumption. The LQLC estimator is less affected by these observations because the logistic model tolerates large residuals more naturally and the IRLS solution down-weights them according to their magnitudes.\par

The quantitative results in Table \ref{tab: light urban} confirm this pattern. LQLC reduces the 3D RMSE from 34.82\,m to 30.96\,m, corresponding to an 11.06\% improvement, and reduces the 3D STD from 11.13\,m to 4.45\,m, corresponding to a 59.99\% improvement. The light urban dataset therefore establishes the practical feasibility of the logistic-based estimator and already indicates a marked gain in positioning stability.\par

It is also worth noting that LS performs competitively during a small portion of the dataset, particularly in short intervals where the observations appear to be less affected by multipath. This behavior is consistent with the Gaussian model remaining suitable when line-of-sight measurements dominate. Even in such intervals, however, the overall performance of LQLC remains comparable, indicating that the logistic formulation improves robustness without imposing a substantial penalty under relatively benign urban conditions.\par

\subsubsection{Medium urban}

\begin{figure}[htb!]
    \addsubFig{0.44}{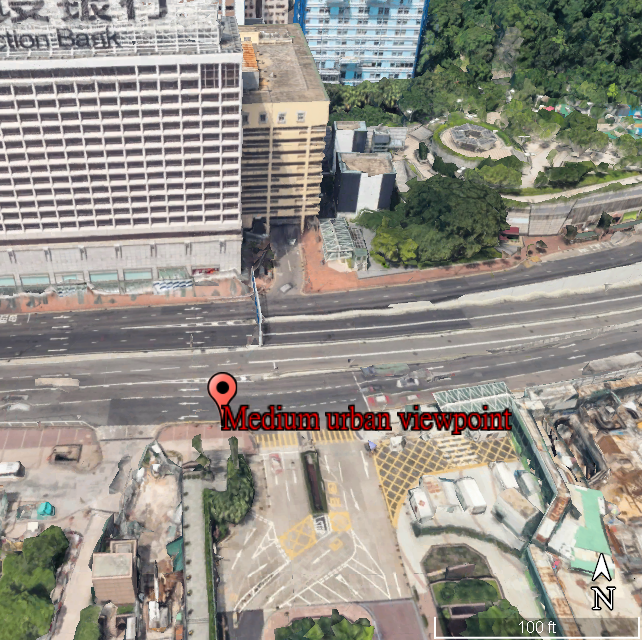}{}
    \addsubFig{0.49}{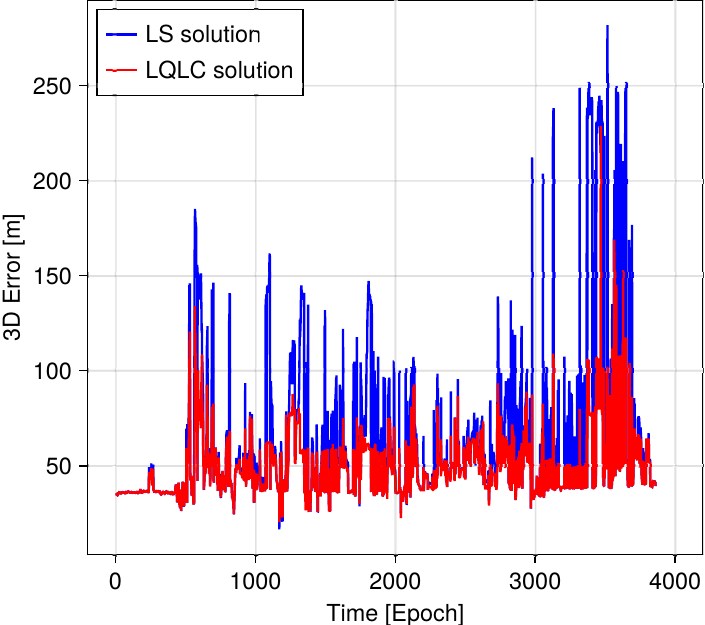}{}
    \caption{Analysis of the medium urban dataset collected at the LLH position of (22.294614, 114.173417, 3.0m): (a) Google Earth snapshot of the receiver's surroundings; (b) 3D positioning errors given by LS and LQLC estimators}
    \label{fig: medium urban}
\end{figure}

Figure \ref{fig: medium urban}a shows that the medium urban site is located beside a high-rise building and dense vegetation. Relative to the light urban case, this environment forms a partial urban canyon and produces more persistent diffraction and reflection effects. Figure \ref{fig: medium urban}b shows that the contrast between LS and LQLC becomes more pronounced in this setting. LS now exhibits frequent error spikes above 200\,m, with a peak close to 280\,m near epoch 3500. The LQLC solution remains more stable and limits most spikes to below 150\,m. The stronger separation between the two estimators is consistent with the expectation that the benefit of a heavier-tailed error model becomes more visible as contaminated measurements occur more frequently.\par 

Table \ref{tab: medium urban} shows that these qualitative differences translate into substantial metric improvements. The 3D RMSE is reduced from 74.71\,m to 51.84\,m, a 30.61\% reduction, while the 3D STD decreases from 42.12\,m to 16.49\,m, a 60.84\% reduction. Compared with the light urban case, the medium urban results more clearly demonstrate the advantage of the logistic-based estimator as the level of environmental degradation increases. Notably, the STD reduction remains close to 60\% in both the light and medium urban datasets, suggesting that the main advantage of LQLC lies not only in reducing average error, but also in suppressing the large and irregular excursions that make urban positioning unreliable in practice.\par

\begin{table}[htb!]
    \caption{Comparison between LS and LQLC positioning solutions in the medium urban environment.}
    \label{tab: medium urban}
    \begin{tblr}
            {colspec={X[c]X[c]X[c]X[c]},
            width=\textwidth,
            row{even} = {white,font=\small},
            row{odd} = {bg=black!10,font=\small},
            row{1} = {bg=black!20,font=\bfseries\small},
            hline{Z} = {1pt,solid,black!60},
            rowsep=3pt
            }
            \textbf{Metric}& \textbf{LS solutions} & \textbf{LQLC solutions} & \textbf{Percentage of reduction}  \\
            3D RMSE & 74.71m & 51.84m & 30.61\% \\
            3D STD & 42.12m & 16.49m & 60.84\%
    \end{tblr}
\end{table}

\subsubsection{Deep urban}

\begin{figure}[htb!]
    \addsubFig{0.44}{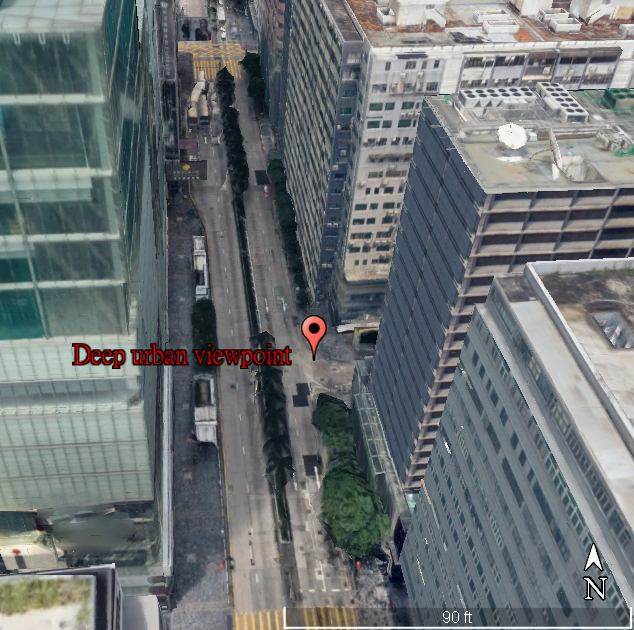}{}
    \addsubFig{0.49}{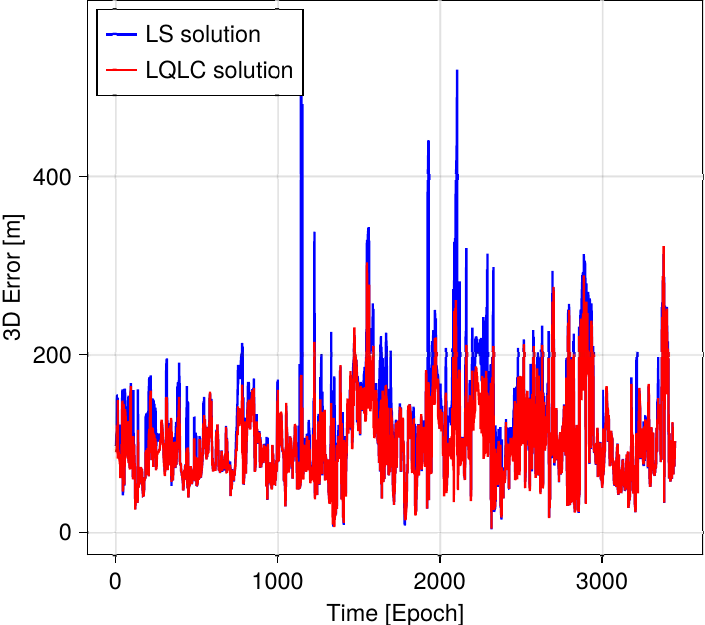}{}
    \caption{Analysis of the deep urban dataset collected at the LLH position of (22.296989, 114.17230, 3.0m): (a) Google Earth snapshot of the receiver's surroundings; (b) 3D positioning errors given by LS and LQLC estimators}
    \label{fig: deep urban}
\end{figure}

Figure \ref{fig: deep urban}a shows that the deep urban site is surrounded by high-rise buildings and represents the most challenging environment among the three datasets. Severe signal blockage, NLOS reception, and strong multipath contamination are all prevalent in this scenario. Figure \ref{fig: deep urban}b shows that both estimators degrade substantially under these conditions, with frequent error spikes exceeding 200\,m. The LS solution occasionally exceeds 400\,m, whereas the LQLC solution generally remains lower and limits the largest excursions to about 300\,m. The benefit of LQLC is therefore still visible, although the margin between the two methods narrows compared with the light and medium urban cases.\par

The quantitative comparison in Table \ref{tab: deep urban} confirms this interpretation. LQLC reduces the 3D RMSE from 131.92\,m to 112.85\,m, corresponding to a 14.46\% improvement, and reduces the 3D STD from 62.16\,m to 45.32\,m, corresponding to a 27.10\% improvement. These gains are smaller than those in the light and medium urban datasets, indicating that very severe urban error processes challenge both Gaussian and logistic models. Nevertheless, LQLC continues to outperform LS in both accuracy and precision across all three urban scenarios.\par

A possible reason for the reduced performance is that, in a deep urban canyon, the frequency and severity of corrupted measurements can exceed the tail behavior anticipated by both the Gaussian and logistic models. Under such conditions, no simple single-distribution model can fully absorb the underlying propagation complexity. The deep urban results therefore suggest both the practical value of LQLC and the remaining difficulty of urban GNSS positioning under extreme signal degradation.\par

\begin{table}[htb!]
    \caption{Comparison between LS and LQLC positioning solutions in the deep urban environment.}
    \label{tab: deep urban}
    \begin{tblr}
            {colspec={X[c]X[c]X[c]X[c]},
            width=\textwidth,
            row{even} = {white,font=\small},
            row{odd} = {bg=black!10,font=\small},
            row{1} = {bg=black!20,font=\bfseries\small},
            hline{Z} = {1pt,solid,black!60},
            rowsep=3pt
            }
            \textbf{Metric}& \textbf{LS solutions} & \textbf{LQLC solutions} & \textbf{Percentage of reduction}  \\
            3D RMSE & 131.92m & 112.85m & 14.46\% \\
            3D STD & 62.16m & 45.32m & 27.10\%
    \end{tblr}
\end{table}

\subsection{Computation efficiency analysis}
The computational cost of the proposed LQLC estimator was compared with that of LS on an Intel Core i7-1355U processor. For each urban dataset, one fixed positioning instance was repeatedly processed in 1000 Monte-Carlo runs so that both the convergence iterations and the execution times could be examined under the same operating condition. \par

\begin{figure}[htb!]
    \centering
    \addsubFig{0.31}{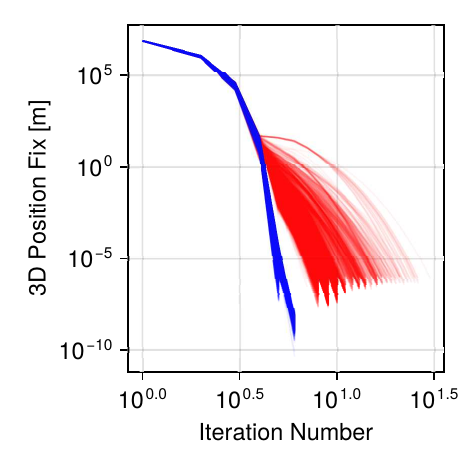}{}
    \addsubFig{0.31}{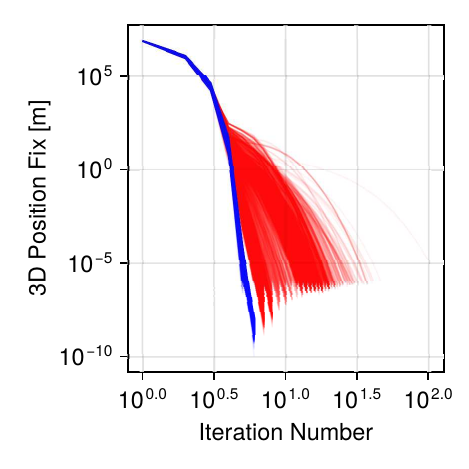}{}
    \addsubFig{0.31}{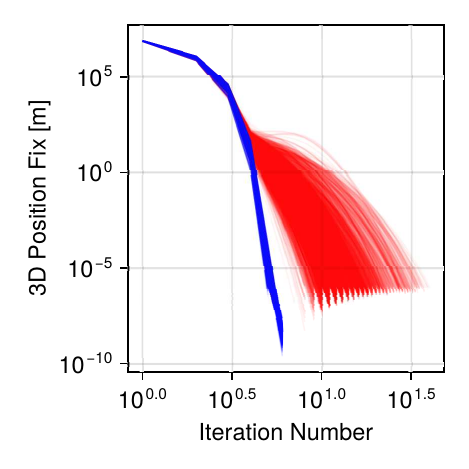}{}
    \caption{Iteration counts required for convergence over 1000 repeated runs of one fixed positioning instance, comparing LS (blue) and LQLC (red) in the (a) light, (b) medium, and (c) deep urban datasets.}
    \label{fig: convergence v.s. iteration for 3}
\end{figure}

Figure \ref{fig: convergence v.s. iteration for 3} shows the iteration counts required for convergence of each estimator. The LS estimator converges more rapidly in terms of iteration count. Across the three urban datasets, the LS solution typically converges at approximately $10^{0.7}$ iterations, whereas LQLC generally requires about $10^{1.5}$ iterations. Since both estimators use the IRLS solver, this difference is attributed to the error-model-induced weighting behavior. For LS, the Gaussian error model leads to a constant weighting function, so the iterative update remains comparatively simple once the linearized model is formed. For LQLC, however, the logistic error model leads to residual-dependent weights that must be adjusted as the current residuals change. As a result, the solver needs more iterations to settle to a stable convergence point. Even so, the increase remains moderate rather than excessive. Even in the deep urban dataset, the required number of iterations stays within a limited range, indicating that the convex QLC loss preserves stable convergence behavior despite the more involved weighting process.\par

\begin{figure}[htb!]
    \centering
    \addsubFig{0.31}{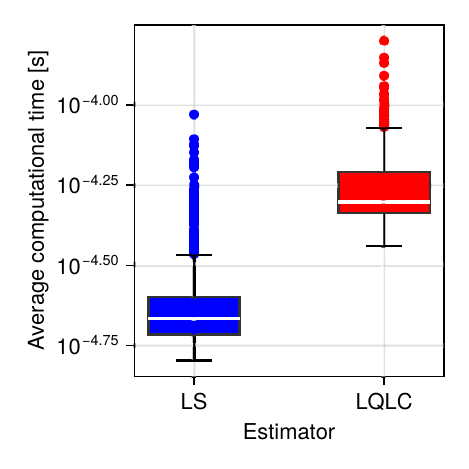}{}
    \addsubFig{0.31}{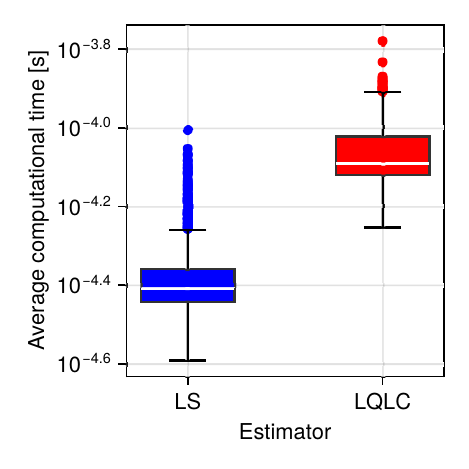}{}
    \addsubFig{0.31}{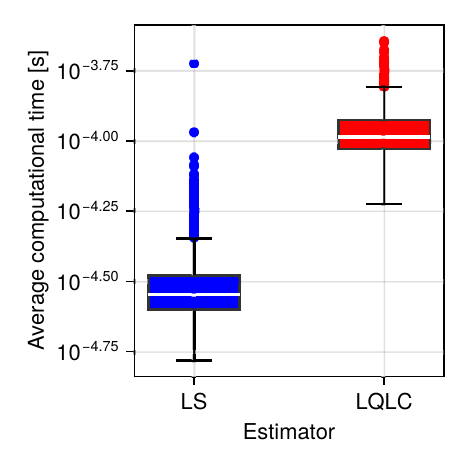}{}
    \caption{Boxplots of the average computational time over 1000 repeated runs of one fixed positioning instance, comparing LS and LQLC in the (a) light, (b) medium, and (c) deep urban datasets. The whiskers extend to $1.5\times$ IQR.}
    \label{fig: average computation time for 3}
\end{figure}

The practical consequence of this iteration gap is shown in Figure \ref{fig: average computation time for 3}, which plots the average computational time of each estimator. As expected, the mean runtime of LQLC is consistently higher than that of LS in all three urban datasets. However, the difference remains small in absolute terms. The average computational time of both estimators stays below $2.5\times 10^{-4}$\,s per epoch, which is fully compatible with real-time single-epoch positioning. Therefore, although the logistic-based estimator requires more iterations, the resulting computational burden remains modest from an operational perspective.\par

Figure \ref{fig: average computation time for 3} also reveals a difference between the two estimators in the dispersion of the runtime statistics. The boxplots show that the standard deviation of the LQLC execution time is slightly smaller than that of LS. A plausible explanation is that, although LQLC requires more iterations on average, its residual-dependent weighting suppresses the influence of extreme observations and makes the iterative trajectory more repeatable over the 1000 runs. By contrast, the constant weighting function used by LS does not attenuate large residuals in the same way. When the fixed positioning instance contains stronger outlying effects, the corresponding updates can be perturbed more noticeably, which increases the spread of the LS runtime distribution. This behavior suggests that LQLC not only remains computationally feasible on average, but also exhibits relatively stable computational performance in challenging urban environments.

\subsection{Controlled scale-mismatch sensitivity analysis}
The real-data experiments above show that LQLC improves urban positioning performance in light, medium, and deep urban conditions. To complement that evidence, a controlled Monte Carlo analysis was conducted to examine the sensitivity of LQLC to mismatch between the true logistic scale and the scale used in estimation. A fixed satellite geometry extracted from one real epoch in the medium urban dataset was used in this simulation, with only the GPS and Beidou observations retained to match the real-data experiment setting. The geometry matrix was formed by linearizing the pseudorange model at the known receiver reference position of that dataset. The pseudorange errors were generated from a zero-mean logistic distribution with a fixed true scale, and the assumed scale $s_{\mathrm{assumed}}$ in LQLC was varied through the mismatch ratio $\alpha=s_{\mathrm{assumed}}/s_{\mathrm{true}}$ from 0.1 to 10.\par

\begin{figure}[htb!]
    \centering
    \addsubFig{0.4}{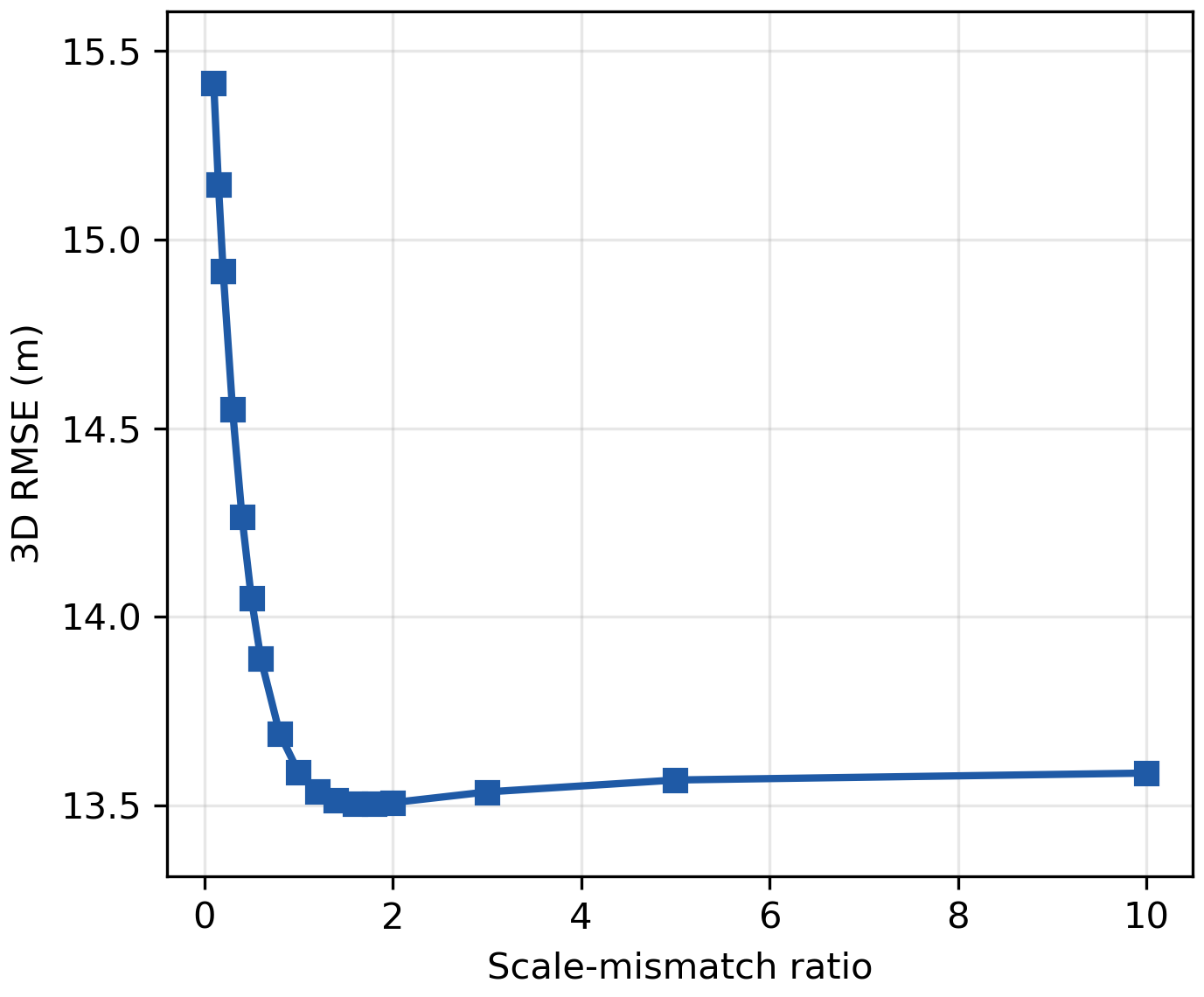}{}
    \addsubFig{0.4}{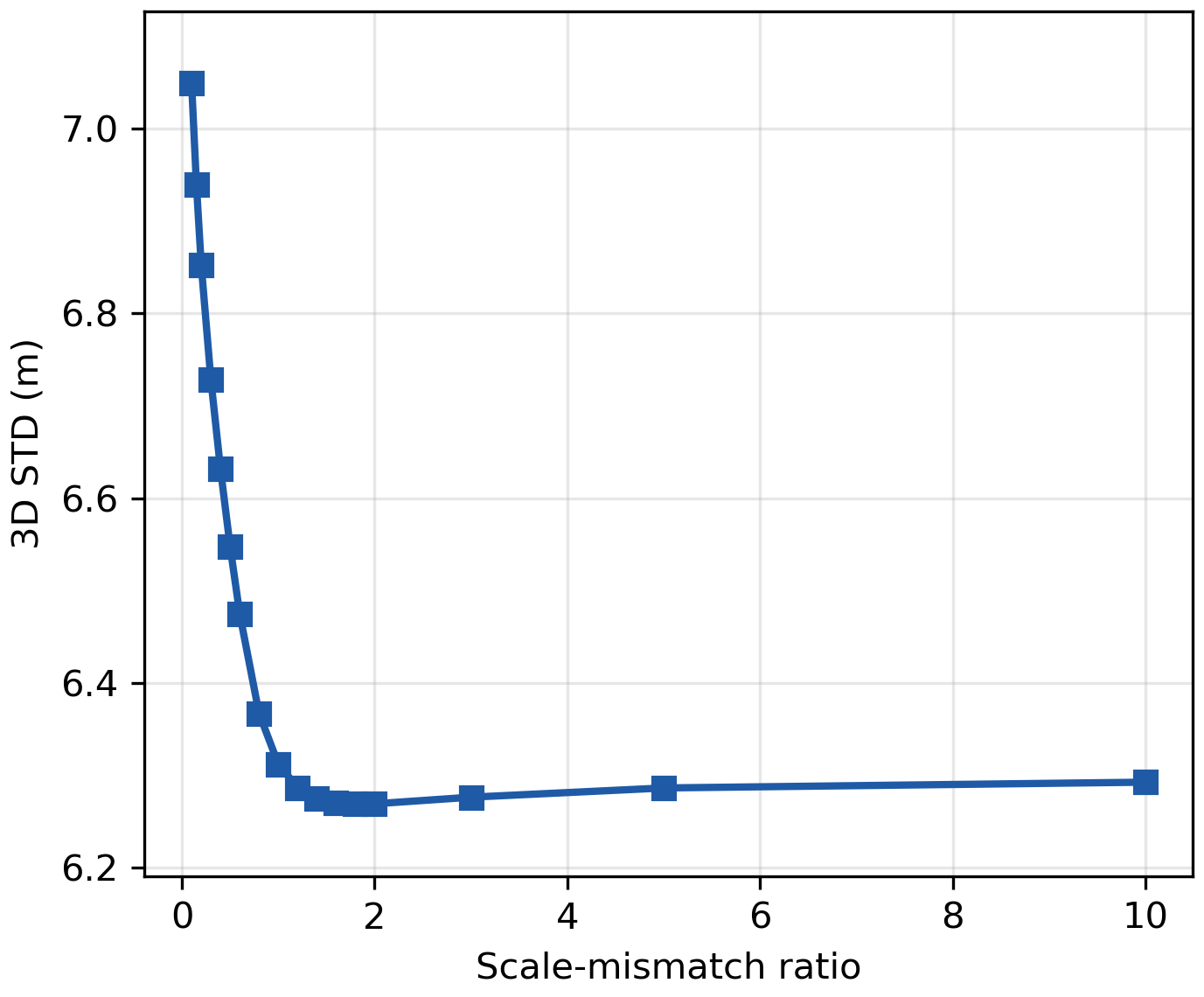}{}
    \caption{Controlled scale-mismatch sensitivity analysis for LQLC under a fixed GPS-and-Beidou geometry extracted from one medium-urban real-data epoch. Panel (a) shows 3D RMSE and panel (b) shows 3D STD as the assumed logistic scale is varied through the mismatch ratio $\alpha=s_{\mathrm{assumed}}/s_{\mathrm{true}}$.}
    \label{fig: scale mismatch sensitivity}
\end{figure}

Figure \ref{fig: scale mismatch sensitivity} shows a clear asymmetry in the scale-mismatch sensitivity of LQLC. When the assumed scale is severely underestimated, the positioning performance deteriorates rapidly. The 3D RMSE decreases from 15.41\,m at $\alpha=0.1$ to about 13.50--13.51\,m near $\alpha=1.6$--$1.8$, while the 3D STD decreases from 7.05\,m to about 6.27\,m over the same range. By contrast, once the assumed scale becomes moderately large, the performance changes only gradually. Even when $\alpha$ is increased to 10, the 3D RMSE and 3D STD remain at 13.59\,m and 6.29\,m, respectively. This figure therefore shows that LQLC is much more sensitive to severe underestimation of the logistic scale than to overestimation.\par

This behavior can be explained through the normalized residual $\bar{r}=r/s$ and the induced weighting function of LQLC. When the assumed scale is too small, the normalized residual becomes excessively large even for measurements with only moderate raw residuals. As a result, many observations enter the saturated downweighting regime too early, and useful measurement information is suppressed more aggressively than intended. This is why small scale values lead to both larger positioning errors and substantially more IRLS iterations. When the assumed scale becomes larger, however, the weighting function varies more gradually with the residual, so further scale enlargement changes the estimator more mildly. The practical issue is therefore not exact scale matching, but avoiding severe underestimation of the logistic scale. This analysis indicates that the advantage of LQLC does not rely on an unrealistically precise estimate of the fitted logistic scale, but it does depend on not choosing an overly small scale that over-penalizes the residuals.\par

\section{Conclusion and future work}\label{sec:conclusion}
This study presented a logistic-based formulation for urban GNSS positioning. Motivated by the heavy-tailed behavior of urban pseudorange errors, the corresponding maximum likelihood estimator, the LQLC estimator, was derived and shown to be solvable efficiently using IRLS. The resulting framework provides a principled alternative to the conventional Gaussian-based LS estimator under urban positioning conditions.\par

Experiments in light, medium, and deep urban environments showed that LQLC consistently improves positioning performance relative to LS. Across the three evaluated scenarios, the 3D RMSE reduction ranged from approximately 11\% to 31\%, while the 3D STD reduction ranged from approximately 27\% to 61\%. The computation-efficiency analysis showed that the proposed method remains compatible with real-time positioning requirements. The controlled scale-mismatch analysis further showed that the sensitivity of LQLC to scale mismatch is asymmetric: severe underestimation is clearly more harmful than overestimation, whereas the estimator remains effective over a fairly wide range on the larger-scale side. This result strengthens the practical case for logistic error modeling in urban GNSS positioning, where the fitted scale parameter is estimated from finite data and cannot be expected to be exact.\par

Future work will focus on two directions. First, the logistic model may support the development of reliable and computationally efficient test statistics to complement conventional Gaussian-based receiver autonomous integrity monitoring (RAIM). Second, the same modeling framework may be extended to other GNSS state-estimation problems, including differential positioning and multi-sensor fusion.

\appendix
\makeatletter
\renewcommand{\@seccntformat}[1]{Appendix \csname the#1\endcsname\quad}
\makeatother

\section{MLE Derivations with a Logistic Error Assumption}
\label{app: logi mle}


\begin{align}
   \hat{\textbf{x}} &= \argmax\ \prod_{i=1}^n f_L(\bm{\varepsilon}_i;0,s_i) \\
   &= \argmax\ \prt{\prod_{i=1}^n \frac{1}{s_i}} \prt{\prod_{i=1}^n  \frac{1}{ \exp\left(\frac{\textbf{y}_{(i)}-\textbf{H}_{(i,:)} \textbf{x}}{s_i}\right)+\exp\left(-\frac{\textbf{y}_{(i)}-\textbf{H}_{(i,:)} \textbf{x}}{s_i}\right) +2}}\\
     &= \argmin \sum_{i=1}^n \ln\prt{s_i}+ \sum_{i=1}^n \ln\prt{2 \cosh \prt{\frac{\textbf{y}_{(i)}-\textbf{H}_{(i,:)} \textbf{x}}{s_i}}+2} \\
    &=  \argmin \sum_{i=1}^n \ln \prt{\cosh \prt{\frac{\textbf{y}_{(i)}-\textbf{H}_{(i,:)} \textbf{x}}{s_i}}+1}
\end{align}

\section{MLE Derivations with a Gaussian Error Assumption}
\label{app: gaussian mle}
Based on Gaussian error assumptions, where each error source follows $\mathcal{N}(0, \sigma_i)$, the likelihood function is given by the product of individual Gaussian PDFs. The MLE seeks to maximize this likelihood, which is equivalent to minimizing the negative log-likelihood:
\begin{equation}
    \hat{\textbf{x}}=\argmin \sum_{i=1}^n -\ln\prt{\frac{1}{\sqrt{2\pi}\sigma_i} e^{-\frac{1}{2}\prt{\frac{\textbf{y}_{(i)}-\textbf{H}_{(i,:)}\textbf{x}}{\sigma_i}}^2}}
\end{equation}
Ignoring constant terms, this simplifies to the least squares estimator:
\begin{equation}
    \hat{\textbf{x}}=\argmin \sum_{i=1}^n  \frac{1}{2}\prt{\frac{\textbf{y}_{(i)}-\textbf{H}_{(i,:)}\textbf{x}}{\sigma_i}}^2=\argmin \sum_{i=1}^n\mathcal{J}_{LS,i}. \label{equ: LS estimator app}
\end{equation}
The solution to the LS estimator is in the form of weighted least squares (WLS):
\begin{equation}
    \hat{\textbf{x}}=\prt{\textbf{H}^{\top}\textbf{W} \textbf{H}}^{-1}\textbf{H}^{\top}\textbf{W} \textbf{y}, \label{equ: WLS solution}
\end{equation}
where the weighting matrix $\textbf{W}$ is a diagonal matrix with $\textbf{W}_{(i,i)} = \frac{1}{\sigma_i^2}$.

\section{Formulation of BGMM and student's t distribution}
\label{app: BGMM student formulation}
BGMM is a weighted mixture of two Gaussian components and its probability density function (PDF) gives that 
\begin{equation}
    f_{BGMM}(x; p_1, \mu_1, \sigma_1, \mu_2, \sigma_2) = p_1 f_N(x; \mu_1, \sigma_1) + (1-p_1)f_N(x; \mu_2, \sigma_2). 
\end{equation}
In this expression, $p_1$, $\mu_1$, and $\sigma_1$ are the weighting, location, and scale parameters for the first Gaussian component, while $\mu_2$ and $\sigma_2$ are the location and scale parameter for the second component.\par
The PDF of a generalized Student's t distribution is given by
\begin{equation}
    f_{t}(x; \mu, \lambda, \nu) = \frac{1}{\lambda} \cdot \frac{\Gamma\prt{\frac{\nu+1}{2}}}{\sqrt{\nu\pi}\Gamma\prt{\frac{\nu}{2}}} \cdot \prt{1+\frac{\prt{\frac{x-\mu}{\lambda}}^2}{\nu}}^{-\frac{\nu+1}{2}},
\end{equation}
where $\mu$ and $\lambda$ are the location and scale parameter, while $\nu$ denotes the degree of freedom. \par

\printbibliography

\end{document}